\pdfoutput=1
\documentclass[11pt]{article}
\usepackage[utf8]{inputenc}
\usepackage[top=70pt,bottom=70pt,left=60pt,right=60pt]{geometry}
\usepackage{amsmath}
\usepackage[fleqn]{mathtools}
\usepackage{hyperref}
\usepackage{amssymb}
\usepackage{graphicx,caption,subcaption}
\usepackage{cite}
\usepackage{bm}
\usepackage{ mathrsfs }
\usepackage[space]{grffile}
\usepackage{graphicx}
\newcommand{\overbar}[1]{\mkern 1.5mu\overline{\mkern-1.5mu#1\mkern-1.5mu}\mkern 1.5mu}

\begin{document}
\thispagestyle{empty}

{\hbox to\hsize{
\vbox{\noindent March 2020 (revised)}}}

\noindent
\vskip2.0cm
\begin{center}

{\Large\bf Volkov--Akulov--Starobinsky supergravity revisited}

\vglue.3in

Yermek Aldabergenov
\vglue.2in

{\it Department of Physics, Faculty of Science, Chulalongkorn University,}\\
{\it Thanon Phayathai, Pathumwan, Bangkok 10330, Thailand}\\
and\\
{\it Institute of Experimental and Theoretical Physics, Al-Farabi Kazakh National University, }\\
{\it 71 Al-Farabi Avenue, Almaty 050040, Kazakhstan}\\
\vglue.1in
yermek.a@chula.ac.th
\end{center}

\vglue.3in

\begin{center}
{\Large\bf Abstract}
\vglue.2in
\end{center}

We find new realizations of Volkov--Akulov--Starobinsky supergravity, i.e. Starobinsky inflationary models in supergravity coupled to a nilpotent superfield describing Volkov--Akulov goldstino. Our constructions are based on the no-scale K\"ahler potential $K=-3\log(T+\overbar{T})$ for the inflaton field, and can describe de Sitter vacuum after inflation where supersymmetry is broken by the goldstino auxiliary component. In fact, we show that a more general class of models with $K=-\alpha\log(T+\overbar{T})$ for $3\leq\alpha\lesssim 6.37$ can accomodate Starobinsky-like inflation with the universal prediction $n_s\simeq 1-\frac{2}{N_e}$ and $r\simeq \frac{4\alpha}{(\alpha-2)^2N_e^2}$, while for $6.37\lesssim\alpha\lesssim 7.23$ viable hilltop inflation is possible (with $n_s$ and $r$ close to the above expressions). We derive the full component action and the masses of sinflaton, gravitino, and inflatino that are generally around the inflationary Hubble scale. Finally, we show that one of our models can be dualized into higher-derivative supergravity with constrained chiral curvature superfield.

\newpage

\section*{Introduction}

Nilpotent superfields have proved to be an invaluable tool for phenomenological supergravity: they can be used for de Sitter uplifting without scalar fields \cite{Bergshoeff:2015tra,Hasegawa:2015bza,Kallosh:2015sea,Kallosh:2015tea}, inflationary model building \cite{Antoniadis:2014oya,Kallosh:2014hxa,Kallosh:2015lwa,Ferrara:2014kva,DallAgata:2014qsj,Hasegawa:2015era,Delacretaz:2016nhw,Argurio:2017joe,Dalianis:2017okk}, and describing string low-energy effective theories in a manifestly supersymmetric way \cite{Kallosh:2014wsa,Bergshoeff:2015jxa,Kallosh:2015nia,Bandos:2015xnf,Garcia-Etxebarria:2015lif,Aparicio:2015psl,Kallosh:2016aep,Cribiori:2019hod}.

The usefulness of nilpotent (chiral) superfields in the context of inflationary model building stems from the fact that once the nilpotency constraint,
\begin{equation}
    \mathbf{S}^2=0~,\label{S_nil}
\end{equation}
is imposed on the superfield $\mathbf{S}$ (we use boldface letters for superfields, and the same non-bold letters for their leading components), its leading, scalar component $S$ is replaced by the fermion bilinear $\sim(\chi^s)^2$ and vanishes from the scalar potential. More specifically, consider the scalar chiral superfield that can be expanded as (using the notations and conventions of Ref. \cite{Wess:1992cp})
\begin{equation}
    \mathbf{S}=S+\sqrt{2}\Theta\chi^s+\Theta^2 F^s~,
\end{equation}
where $\chi^s$ is its chiral fermion, and $F^s$ is its auxiliary component. It can easily be checked that the nilpotency constraint \eqref{S_nil} is solved by $S=(\chi^s)^2/(2F^s)$. This implies that $F^s$ must be non-vanishing and the construction features spontaneously broken $N=1$ supersymmetry that is non-linearly realized on the goldstino $\chi^s$ \cite{Rocek:1978nb,Lindstrom:1979kq,Ivanov:1978mx,Casalbuoni:1988xh,Komargodski:2009rz}. It was shown in Ref. \cite{Kuzenko:2010ef} that the resulting action is equivalent (via a non-linear field redefinition) to the original Volkov-Akulov (VA) action \cite{Volkov:1973ix}.~\footnote{Recently an alternative approach (without nilpotent superfields) to Volkov--Akulov supergravity in de Sitter space was proposed, that uses unimodular supergravity \cite{Nagy:2019ywi,Anero:2019ldx}.}

So, on the one hand, nilpotent superfields add flexibility of the multi-superfield inflationary models, and on the other, spontaneously break supersymmetry -- all of this without introducing extra dynamical scalars (the corresponding scalars are assumed to be decoupled from low-energy theories \cite{Farakos:2013ih}).

In this study we will be focusing on the Starobinsky(-like) inflation \cite{Starobinsky:1980te}, motivated by its remarkable agreement with CMB measurements \cite{Akrami:2018odb}. In Ref. \cite{Cecotti:1987sa} it was shown by Cecotti, that (old-minimal) $R+R^2$ supergravity is dual to the standard supergravity coupled to two chiral multiplets with
\begin{equation}
    K=-3\log(T+\overbar{T}-C\overbar{C})~,~~~W=\gamma C(T-1/2)~,\label{KW_Cecotti}
\end{equation}
where $T=\mathbf{T}|_{\Theta=0}$ and $C=\mathbf{C}|_{\Theta=0}$ are the two chiral scalars, and $\gamma$ is some constant (throughout the paper we will use Planck units, $M_P=1$). For $C=0$ this leads to the Starobinsky scalar potential for appropriately normalized real part of $T$.~\footnote{Although the original potential has tachyonic instability at $C=0$, it can be removed by adding quartic correction $\sim|C|^4$ to the K\"ahler potential \eqref{KW_Cecotti}, as was shown in Ref. \cite{Kallosh:2013xya}.} In Ref. \cite{Antoniadis:2014oya} the authors made a first step towards bringing together Starobinsky inflation and Volkov-Akulov supergravity, by replacing the unconstrained superfield $\mathbf{C}$ in the Cecotti model with the nilpotent one $\mathbf{S}$ (see also Ref. \cite{Ozkan:2014cua} for $R^n$-extension of Starobinsky supergravity with nilpotent goldstino). We will refer to the construction of Ref. \cite{Antoniadis:2014oya} as the Antoniadis--Dudas--Ferrara--Sagnotti (ADFS) model. In this model the nilpotency constraint \eqref{S_nil} ensures that the scalar $S$ is replaced by the goldstino bilinear, and the scalar sector includes only the inflaton -- given by ${\rm Re}T$ -- and its superpartner (sinflaton) ${\rm Im}T$ that is heavy during (and after) inflation. There is however one issue that has to be addressed before proceeding to a more realistic setup with matter fields included. At the minimum of the potential of the ADFS model, the auxiliary component of the goldstino vanishes, $\langle F^s\rangle=0$, which renders the solution $S=(\chi^s)^2/(2F^s)$ to the constraint \eqref{S_nil} singular, as was pointed out in Ref. \cite{DallAgata:2014qsj}.~\footnote{In Ref. \cite{DallAgata:2014qsj} the authors also propose a different class of Volkov--Akulov--Starobinsky supergravity models where the K\"ahler potential has the simplest shift-symmetric form, $K=(T+\overbar{T})^2/2$.} The goal of the present work is to resolve this issue by introducing minimal amount of modifications to the K\"ahler potential and superpotential of the original theory.

In Section 1 we review the ADFS model and discuss the problem of vanishing $F^s$ in more detail. In Section 2 we show how the issue can be resolved by modifying the model in two different ways, while keeping the no-scale structure of the K\"ahler potential, $K=-3\log(T+\overbar{T}+\ldots)$. Section 3 is devoted to generalization of the K\"ahler potentials of the aforementioned models as $K=-\alpha\log(T+\overbar{T}+\ldots)$ and derivation of the scalar potential that includes a Starobinsky-like inflationary plateau. The full action, including fermions, is derived in Section 4, where we compare masses of the fields at different $\alpha$. In Section 5 we use slow-roll approximation to derive the prediction for the inflationary observables $n_s$ and $r$. In Section 6 we review the gravitational dual of the ADFS model, and show that one of our models can also be dualized into higher-derivative supergravity where the nilpotency constraint for the chiral curvature superfield is modified compared to the ADFS model. Section 7 is left for conclusion, and some basic supergravity formulae and conventions that we use here can be found in Appendix.

\section{The original proposal -- ADFS model}

The ADFS model is based on the following setup
\begin{gather}
    K=-3\log(T+\overbar{T}-S\overbar{S})~,\label{K_ADFS}\\
    W=\lambda+\beta S+\gamma ST~,\label{W_ADFS}
\end{gather}
where $\lambda,\beta,\gamma$ are some real parameters (this superpotential coincides with that of Eq. \eqref{KW_Cecotti} if we set $\beta=-\gamma/2$ and $\lambda=0$), $T$ includes inflaton and sinflaton fields, and $S$ is the leading component of the nilpotent superfield so that $\mathbf{S}^2=S^2=0$. Thus, the K\"ahler potential \eqref{K_ADFS} can be expanded as
\begin{equation}
    K=-3\log(T+\overbar{T})+\frac{3S\overbar{S}}{T+\overbar{T}}~.
\end{equation}

Once the action is derived we can apply the solution $S=(\chi^s)^2/(2F^s)$ to the nilpotency constraint, and after using the parametrization
\begin{equation}
    T=\frac{t_0}{2}\left(e^{\sqrt{\frac{2}{3}}\varphi}+i\sqrt{\frac{2}{3}}\tau\right)~,\label{T_3}
\end{equation}
where $t_0>0$ (i.e. choosing upper-half-plane of the Poincar\'e disk) is the VEV of $T$ so that at the minimum $\varphi=0$, the bosonic Lagrangian reads
\begin{equation}
    e^{-1}{\cal L}=\frac{1}{2}R-\frac{1}{2}(\partial_m\varphi)^2-\frac{1}{2}e^{-2\sqrt{\frac{2}{3}}\varphi}(\partial_m\tau)^2-\frac{\gamma^2}{12}\left(1-e^{-\sqrt{\frac{2}{3}}\varphi}\right)^2-\frac{\gamma^2}{18}e^{-2\sqrt{\frac{2}{3}}\varphi}\tau^2~,\label{L_ADFS}
\end{equation}
where we used $t_0=-2\beta/\gamma$ (found by solving the vacuum equations), assuming that $\beta\gamma<0$ as required for the existence of a stable minimum. The masses of the inflaton $\varphi$ and sinflaton $\tau$ (w.r.t. the Minkowski minimum at $\varphi=0$) are
\begin{equation}
    m_\varphi=m_\tau=\gamma/3~.
\end{equation}
During inflation, $\varphi\gg 1$, the $\tau$ effective mass is unchanged because its kinetic term and mass term are coupled to the same exponential of $\varphi$ and canonical rescaling of $\tau$ fully absorbs any background value of $\varphi$. On the other hand the Hubble scale during inflation is $H\simeq\sqrt{V_{\rm inf}/3}\simeq\gamma/6$, so that $m_\tau\simeq 2H$.

Once the inflaton settles at the minimum $\varphi=0$, we have~\footnote{The relation between $D_iW\equiv W_i+K_iW$ and $F^i$ is given by Eq. \eqref{F_DW} in the Appendix.}
\begin{equation}
    \langle D_TW\rangle=\frac{3\gamma\lambda}{2\beta}~,~~~\langle D_SW\rangle=0~,
\end{equation}
which means that $S=(\chi^s)^2/(2F^s)$ diverges, and the nilpotency constraint is no longer valid. Moreover, SUSY becomes broken by $F^t$ instead of $F^s$. Although we can set $\lambda=0$ so that $F^t$ vanishes, the gravitino mass,
\begin{equation}
    m_{3/2}=\langle e^{K/2}|W|\rangle=\frac{\gamma^{3/2}\lambda}{2\sqrt{2}\beta^{3/2}}~,
\end{equation}
will vanish as well.

\section{Improved models}

Here we will show that adding a single, $T$-linear term in the superpotential can improve upon the original ADFS model by changing the auxiliary VEVs as
\begin{equation}
    \langle D_T W\rangle=0~,~~~\langle D_S W\rangle\neq 0~,
\end{equation}
and introducing a tunable cosmological constant that can be used to describe the dark energy.

Consider the case
\begin{gather}
    K=-3\log(T+\overbar{T}-S\overbar{S})~,\label{K_3I}\\
    W=\lambda-\mu T+\beta S+\gamma ST~,\label{W_3I}
\end{gather}
where we assume that all the parameters $\{\lambda,\mu,\beta,\gamma\}$ are real and non-vanishing. Ignoring the sinflaton for a moment, this leads to the scalar potential (after using the parametrization \eqref{T_3})
\begin{equation}
    V=\frac{\gamma^2}{12}+\frac{1}{3}\left(\beta\gamma-2\mu^2\right)t^{-1}+\frac{1}{3}\left(\beta^2+6\lambda\mu\right)t^{-2}~,\label{Vpre_3I}
\end{equation}
where for convenience we introduced the notation $t\equiv T+\overbar{T}= t_0e^{\sqrt{2/3}\varphi}$. The vacuum value $t_0$ for the above potential can be easily found as
\begin{equation}
    t_0=-2\frac{\beta^2+6\lambda\mu}{\beta\gamma-2\mu^2}~.\label{t0}
\end{equation}
Now, recall that $D_TW$ must vanish at the minimum (and $D_SW$ must not) in order for the $\mathbf{S}$ to be identified with the goldstino superfield. Deriving $D_TW$ for the setup \eqref{K_3I} and \eqref{W_3I} and assuming $\tau=0$ we have
\begin{equation}
    D_TW=\frac{\mu}{2}-\frac{3\lambda}{t}~.
\end{equation}
Requiring $D_TW$ to vanish at $t=t_0$ leads to $t_0=6\lambda/\mu$, so that $\lambda/\mu$ must be positive. Substituting this into Eq. \eqref{t0} we arrive at the condition
\begin{equation}
    \beta\mu=-3\gamma\lambda~.\label{cond1}
\end{equation}
The cosmological constant can be calculated from Eq. \eqref{Vpre_3I} by using $t_0=6\lambda/\mu$,
\begin{equation}
    V_0=\frac{1}{108\lambda^2}(9\gamma^2\lambda^2-6\lambda\mu^3+\beta^2\mu^2+6\beta\gamma\lambda\mu)~.
\end{equation}
Then, we can use Eq. \eqref{cond1} to eliminate e.g. $\beta$ in the cosmological constant and observe that
\begin{equation}
    V_0=-\frac{\mu^3}{18\lambda}<0~,
\end{equation}
i.e. $V_0$ turns out to be negative as long as none of the parameters of the superpotential is zero. 

By looking at Eq. \eqref{t0} it is clear that if we set $\beta=0$, the condition $t_0=6\lambda/\mu$ (i.e. $\langle D_TW\rangle=0$) is automatically satisfied! Moreover, the cosmological constant becomes
\begin{equation}
    V_0=\frac{\gamma^2}{12}-\frac{\mu^3}{18\lambda}~,\label{V0_3I}
\end{equation}
so that we can fine-tune the parameters to yield $V_0\sim 10^{-120}$.

\subsection{The case \texorpdfstring{$\beta=0$}{Lg}}

Let us now analyze in more detail the model \eqref{K_3I} and \eqref{W_3I} with $\beta=0$. After using Eq. \eqref{T_3} with $t_0=6\lambda/\mu$ and eliminating $\gamma$ in terms of $V_0$, $\mu$ and $\lambda$ via Eq. \eqref{V0_3I}, the bosonic Lagrangian reads~\footnote{This can be identified with the original ADFS Lagrangian if we replace $\mu^3/(18\lambda)\rightarrow \gamma^2/12$ and put $V_0=0$.}
\begin{equation}
    e^{-1}{\cal L}=-\frac{1}{2}(\partial_m\varphi)^2-\frac{1}{2}e^{-2\sqrt{\frac{2}{3}}\varphi}(\partial_m\tau)^2-V_0-\frac{\mu^3}{18\lambda}\left(1-e^{-\sqrt{\frac{2}{3}}\varphi}\right)^2-\frac{\mu^3}{27\lambda}e^{-2\sqrt{\frac{2}{3}}\varphi}\tau^2~,\label{L_3I}
\end{equation}
where the $\tau^2$-term in the scalar potential was originally proportional to $\gamma^2$ that we eliminated via Eq. \eqref{V0_3I} while neglecting $V_0$ due to its relative smallness. In what follows, we will similarly eliminate $\gamma$ via Eq. \eqref{V0_3I} as $\gamma\approx\sqrt{2\mu^3/(3\lambda)}$.

\begin{figure}
\centering
\begin{subfigure}{.47\textwidth}
  \centering
  \includegraphics[width=1\linewidth]{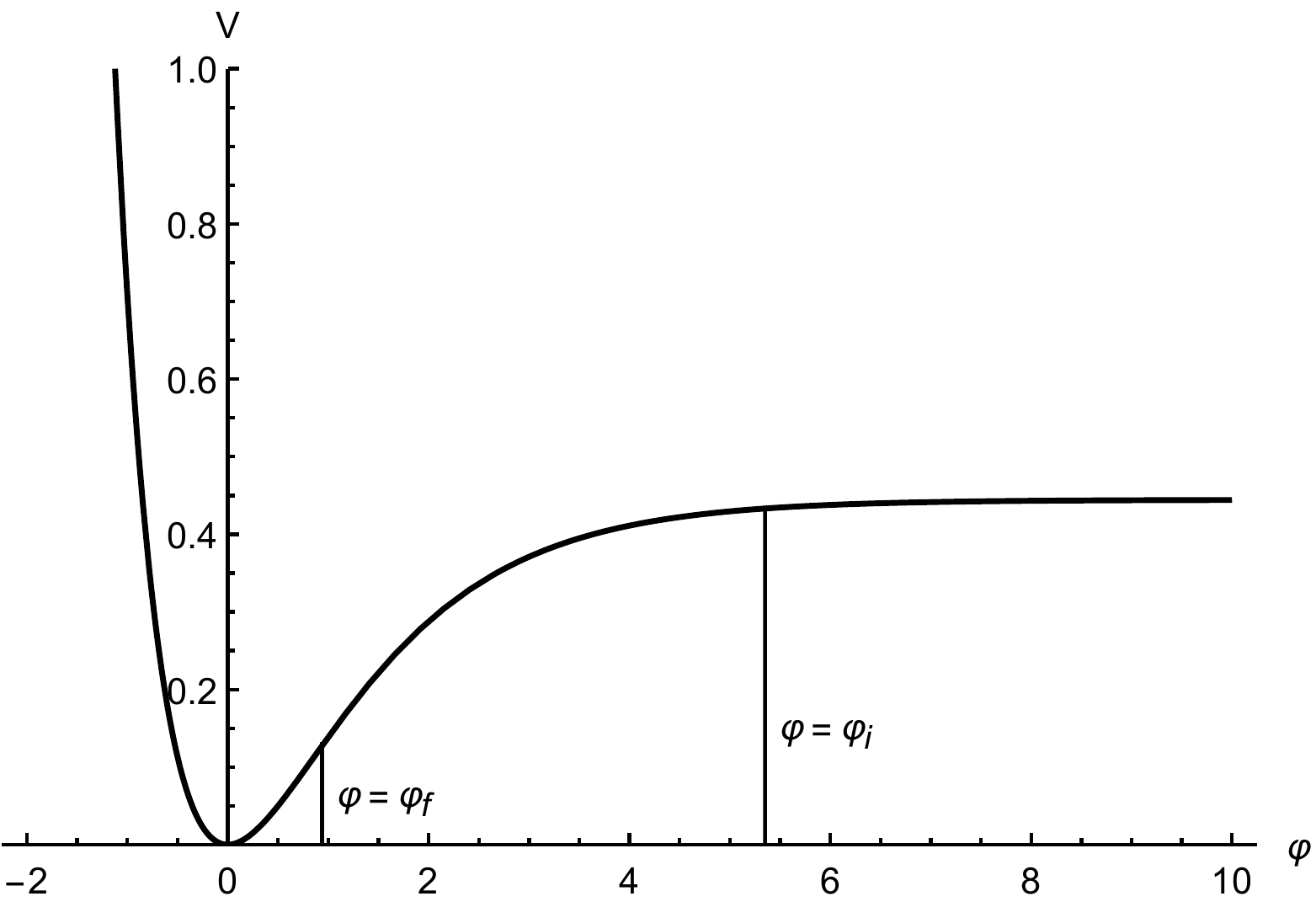}
  \caption{}
  \label{Fig_V_varphi}
\end{subfigure}
\hspace{1em}
\begin{subfigure}{.47\textwidth}
  \centering
  \includegraphics[width=1\linewidth]{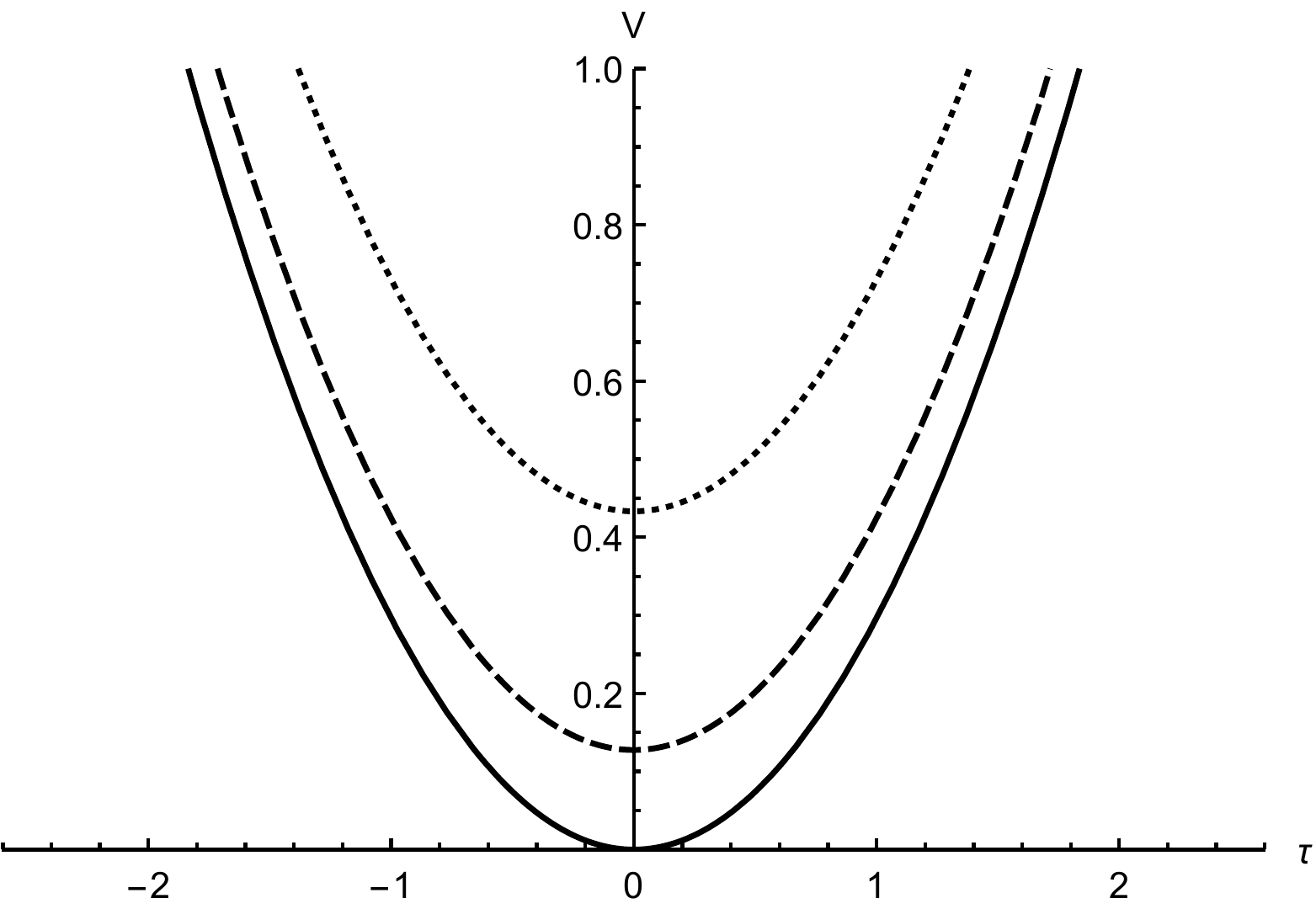}
  \caption{}
  \label{Fig_V_tau}
\end{subfigure}
\captionsetup{width=1\linewidth}
\caption{The scalar potential of \eqref{L_3I} for $\mu=2,\lambda=1$ and $V_0=0$. Subfigure (a) represents $\varphi$-dependent slice ($\tau=0$), while Subfigure (b) represents $\tau$-dependent slice for different background values of $\varphi$: solid line stands for $\varphi=0$, dashed line for $\varphi=\varphi_f$, dotted line for $\varphi=\varphi_i$.}
\label{Fig_V}
\end{figure}

The scalar masses can be read off as $m^2_\varphi=m^2_\tau=2\mu^3/(27\lambda)$. The potential in $\varphi$-direction is presented in Figure \ref{Fig_V_varphi} where we include the points $\varphi_i$ and $\varphi_f$ representing the start and end of (observable) inflation, respectively, assuming $55$ e-foldings. Due to the coupling of $\tau$-kinetic term to the inflaton, we draw the potential in $\tau$-direction separately at different reference points $\varphi=0$, $\varphi_i$, and $\varphi_f$, after canonical rescaling of $\tau$ -- see Figure \ref{Fig_V_tau}. 

As we already mentioned, $\langle D_TW\rangle=\langle F^t\rangle=0$ when substituting $t_0=6\lambda/\mu$, while $\langle D_SW\rangle=3\gamma\lambda/\mu$ and the auxiliary field $\langle \tilde{F}^s\rangle$ reads~\footnote{The commonly used definition $\tilde{F}^s$ -- given by Eq. \eqref{tilde_F_DW} -- is related to the $\Theta^2$-expansion coefficient $F^s$ as $F^s=e^{-K/6}\tilde{F}^s$, as explained in Appendix.}
\begin{equation}
    \langle \tilde{F}^s\rangle=-\gamma\sqrt{\frac{\lambda}{6\mu}}=-\frac{\mu}{3}\neq 0~.\label{Fs_3I}
\end{equation}
Therefore, $\mathbf{S}$ can be consistently identified as a nilpotent goldstino superfield. Since $\langle \tilde{F}^s\rangle$ is controlled by $\mu$, its value is independent of CMB observations, because they -- specifically observations of the amplitude of scalar perturbations \cite{Akrami:2018odb} -- fix only the ratio $\mu^3/\lambda\sim 10^{-8}$ (in Planck units).

The gravitino mass is $m^2_{3/2}=\mu^3/(54\lambda)$, i.e. $m_\varphi=2m_{3/2}$ and the inflaton can perturbatively decay into two gravitini at the reheating stage. We can also relate it to the inflationary Hubble scale $m_{3/2}\simeq H$, where $H\simeq\sqrt{V_{\rm inf}/3}\sim 10^{-5}$.

This model can be dualized into higher-derivative ($R^2$) supergravity with a constrained chiral curvature superfield, as will be shown in Section 6.

\subsection{The case \texorpdfstring{$\gamma=0$}{Lg} with modified K\"ahler potential}

We find that there exists a similar realization of the Starobinsky model \eqref{L_3I}, albeit with some key differences, if we slightly modify the K\"ahler potential as
\begin{equation}
    K=-3\log\left(T+\overbar{T}-\frac{S\overbar{S}}{(T+\overbar{T})^2}\right)=-3\log(T+\overbar{T})+\frac{3S\overbar{S}}{(T+\overbar{T})^3}~,\label{K_3II}
\end{equation}
and in the superpotential set $\beta\neq 0$, $\gamma=0$:
\begin{equation}
    W=\lambda-\mu T+\beta S~.\label{W_3II}
\end{equation}
In this case the scalar potential becomes
\begin{equation}
    V=\frac{\beta^2}{3}-\frac{2\mu^2}{3}t^{-1}+2\lambda\mu t^{-2}~,\label{Vpre_3II}
\end{equation}
where $t=t_0e^{\sqrt{2/3}\varphi}$ as before. This time $\tau$ does not appear in the scalar potential. The potential for $\tau$ can be generated e.g. along the lines of Refs. \cite{Ellis:1984bs,Ellis:2014gxa} where quartic $\sim (T-\overbar{T})^4$ stabilizing terms were considered as modifications of the no-scale K\"ahler potential.

Comparing the potential \eqref{Vpre_3II} with the potential \eqref{Vpre_3I} at $\beta=0$, it is clear that they only differ in their constant terms. Thus, $t_0=6\lambda/\mu$ is also a minimum for the potential \eqref{Vpre_3II}, and $\langle D_TW\rangle=0$, while $\langle D_SW\rangle=\beta\neq 0$ as required.

Taking similar steps as in the previous subsection, we find the cosmological constant
\begin{equation}
    V_0=\frac{\beta^2}{3}-\frac{\mu^3}{18\lambda}~,
\end{equation}
and use this relation to eliminate $\beta$ in terms of $V_0$, $\mu$ and $\lambda$. Then, the scalar potential reads
\begin{equation}
    V=V_0+\frac{\mu^3}{18\lambda}\left(1-e^{-\sqrt{\frac{2}{3}}\varphi}\right)^2~,\label{V_3II}
\end{equation}
while the kinetic terms are the same as in Eq. \eqref{L_3I}.

As for the $\tilde{F}^s$, its vacuum value is
\begin{equation}
    \langle \tilde{F}^s\rangle=-2\sqrt{6}\beta\left(\frac{\lambda}{\mu}\right)^{\frac{3}{2}}=-2\lambda~.
\end{equation}
In contrast with the previous model, this is controlled by $\lambda$ instead of $\mu$.

\section{Generalization}

Here we consider generalization of the K\"ahler potential as
\begin{equation}
    K=-\alpha\log\left(T+\overbar{T}-\frac{S\overbar{S}}{(T+\overbar{T})^{n-1}}\right)=-\alpha\log(T+\overbar{T})+\frac{\alpha S\overbar{S}}{(T+\overbar{T})^n}~,\label{K_general}
\end{equation}
while the superpotential is kept the same,
\begin{equation}
    W=\lambda-\mu T+\beta S+\gamma ST~.\label{W_general}
\end{equation}
$\alpha$ is a positive real number, and $n$ is an arbitrary real number. After imposing the nilpotency constraint, the K\"ahler potential \eqref{K_general} describes $SU(1,1)/U(1)$ scalar manifold with the K\"ahler curvature $R_K=-2/\alpha$.

The scalar potential of this setup at $\tau=0$ reads
\begin{equation}
    V=\frac{\gamma^2}{4\alpha}t^{n+2-\alpha}+\frac{\beta\gamma}{\alpha}t^{n+1-\alpha}+\frac{\beta^2}{\alpha}t^{n-\alpha}+\frac{\alpha^2-7\alpha+4}{4\alpha}\mu^2t^{2-\alpha}-(\alpha-5)\lambda\mu t^{1-\alpha}+(\alpha-3)\lambda^2t^{-\alpha}~.\label{V_general}
\end{equation}
For our analysis we will also use the necessary condition
\begin{equation}
    \langle D_TW\rangle=\frac{1}{2}(\alpha-2)\mu-\frac{\alpha\lambda}{t_0}=0~.\label{cond_general}
\end{equation}

Let us start with the special value $\alpha=2$ for which the first term in Eq. \eqref{cond_general} vanishes identically. This forces $\lambda=0$ and the potential takes the form
\begin{equation}
    V=\frac{t^n}{2}\left(\frac{\gamma}{2}+\beta t^{-1}\right)^2-\frac{3\mu^2}{4}~.
\end{equation}
Stable minimum exists if $\beta\gamma<0$, but it is always an AdS minimum.

When $\alpha<2$, the $t^{2-\alpha}$-term has a positive power of $t$ while the $t^{-\alpha}$-term has a negative power. That means that we cannot have an inflationary plateau approaching a constant positive value unless $\mu$ or $\lambda$ is zero. But if $\mu$ (or $\lambda$) vanishes, Eq. \eqref{cond_general} forces $\lambda$ (or $\mu$) to vanish as well, so $\lambda=\mu=0$. This leads to $m_{3/2}=0$, which is phenomenologically unacceptable.

Next, consider $2<\alpha<3$. Notice that among the last three terms of Eq. \eqref{V_general} the $t^{-\alpha}$-term is negative, and has the largest power of $t^{-1}$, which destabilizes the potential unless $n$ is chosen in such a way that either of the first three terms has $t^{-m}$ with $m\geq\alpha$. On the other hand, the existence of the inflationary plateau with positive height requires the existence of a constant positive term in the above potential. Such a constant term can come from the first, second, or third term if $n=\alpha-2$, $n=\alpha-1$, or $n=\alpha$, respectively. When $n=\alpha-1$ or $n=\alpha$, the first term has a positive power of $t$, which prevents the required flatness of the potential (because negative powers are also present and come from the last three terms). When $n=\alpha-2$, positive powers of $t$ are absent but the (negative) $t^{-\alpha}$-term is left uncompensated, and will destabilize the potential. Thus, we conclude that $\alpha<3$ is unsuitable for our purposes and in what follows assume that $\alpha\geq 3$.

When $\alpha\geq 3$, the last term of Eq. \eqref{V_general} becomes positive or zero. Starobinsky-like structure of the scalar potential can be obtained by the choice (I) $\beta=0$ and $n=\alpha-2$, or (II) $\gamma=0$ and $n=\alpha$, where $\alpha=3$ reproduces the two Starobinsky models that we described in the previous section. 

The potentials for the cases I and II only differ in their constant terms, and share the two critical points
\begin{equation}
    t_{0(1)}=\frac{2\alpha\lambda}{(\alpha-2)\mu}~,~~~t_{0(2)}=\frac{2\alpha(\alpha-3)\lambda}{(\alpha^2-7\alpha+4)\mu}~.\label{crit_general}
\end{equation}
These describe four different types of scalar potentials depending on the parameter ranges. First, if $\lambda\mu>0$ and $3\leq\alpha\leq\alpha_*$ where $\alpha_*\equiv(7+\sqrt{33})/2\approx 6.37$, the $t_{0(1)}$ is a single critical point that is also the minimum. Second, if $\lambda\mu<0$ and $3<\alpha<\alpha_*$, the $t_{0(2)}$ takes up the role of the minimum. The third possibility is $\lambda\mu>0$ and $\alpha>\alpha_*$. Here the two critical points coexist: $t_{0(1)}$ is the minimum, while $t_{0(2)}$ becomes a local maximum. For all other parameter values no critical points exist.

Substituting the two solutions into Eq. \eqref{cond_general} we obtain (for the cases where $t_{0(1)}$ and $t_{0(2)}$ are the minima, respectively)
\begin{equation}
    \langle D_TW\rangle|_{t_{0(1)}}=0~,~~~\langle D_TW\rangle|_{t_{0(2)}}=\frac{\alpha+1}{\alpha-3}\mu~.
\end{equation}
$\langle D_TW\rangle|_{t_{0(2)}}$ can only vanish if $\mu=0$, but this invalidates the critical points \eqref{crit_general}, i.e. the potential does not admit stable (as well as metastable) minima in this case. Therefore, excluding the second possibility where $\lambda\mu<0$ and $t_{0(2)}$ is the minimum, we are left with $\lambda\mu>0$ and $\alpha\geq 3$.

\subsection{The case I: \texorpdfstring{$\beta=0$}{Lg} and \texorpdfstring{$n=\alpha-2$}{Lg}}

Here we consider $\beta=0$ and $n=\alpha-2$ (with $\alpha\geq 3$), that is reflected in the following setup,
\begin{gather}
    K=-\alpha\log\left(T+\overbar{T}-\frac{S\overbar{S}}{(T+\overbar{T})^{\alpha-3}}\right)~,\label{K_I}\\
    W=\lambda-\mu T+\gamma ST~.\label{W_I}
\end{gather}

After using the generalized form of the parametrization \eqref{T_3},
\begin{equation}
    T=\frac{t_0}{2}\left(e^{\sqrt{\frac{2}{\alpha}}\varphi}+i\sqrt{\frac{2}{\alpha}}\tau\right)~,~~~t_0=\frac{2\alpha\lambda}{(\alpha-2)\mu}~,
\end{equation}
and eliminating $\gamma$ in terms of $V_0,\lambda,\mu$,
\begin{equation}
    \frac{\gamma^2}{4\alpha}=V_0+\frac{12(\alpha-2)^{\alpha-2}\mu^\alpha}{(2\alpha)^\alpha\lambda^{\alpha-2}}~,\label{gamma_V0}
\end{equation}
we obtain the final form of the scalar potential,
\begin{multline}
    V=V_0+\frac{(\alpha-2)^{\alpha-2}\mu^\alpha}{(2\alpha)^\alpha\lambda^{\alpha-2}}\left\lbrace12+\alpha(\alpha^2-7\alpha+4)e^{(2-\alpha)\sqrt{\frac{2}{\alpha}}\varphi}+2\alpha(\alpha-2)(5-\alpha)e^{(1-\alpha)\sqrt{\frac{2}{\alpha}}\varphi}+\right.\\\left.+(\alpha-2)^2(\alpha-3)e^{-\alpha\sqrt{\frac{2}{\alpha}}\varphi}+\left[2\alpha(\alpha-3)e^{-\alpha\sqrt{\frac{2}{\alpha}}\varphi}+\frac{24}{\alpha}e^{-2\sqrt{\frac{2}{\alpha}}\varphi}\right]\tau^2\right\rbrace~,\label{V_I}
\end{multline}
where we set $V_0=0$ everywhere except as the cosmological constant. When $\alpha=3$ we obtain exactly the Starobinsky scalar potential \eqref{L_3I}, whereas for $3<\alpha\leq\alpha_*$ the potential is deformed, but it still includes a Starobinsky-like inflationary plateau for $\varphi\gg1$ (we will perform slow-roll analysis in the upcoming sections).

\begin{figure}
\centering
\begin{subfigure}{.47\textwidth}
  \centering
  \includegraphics[width=1\linewidth]{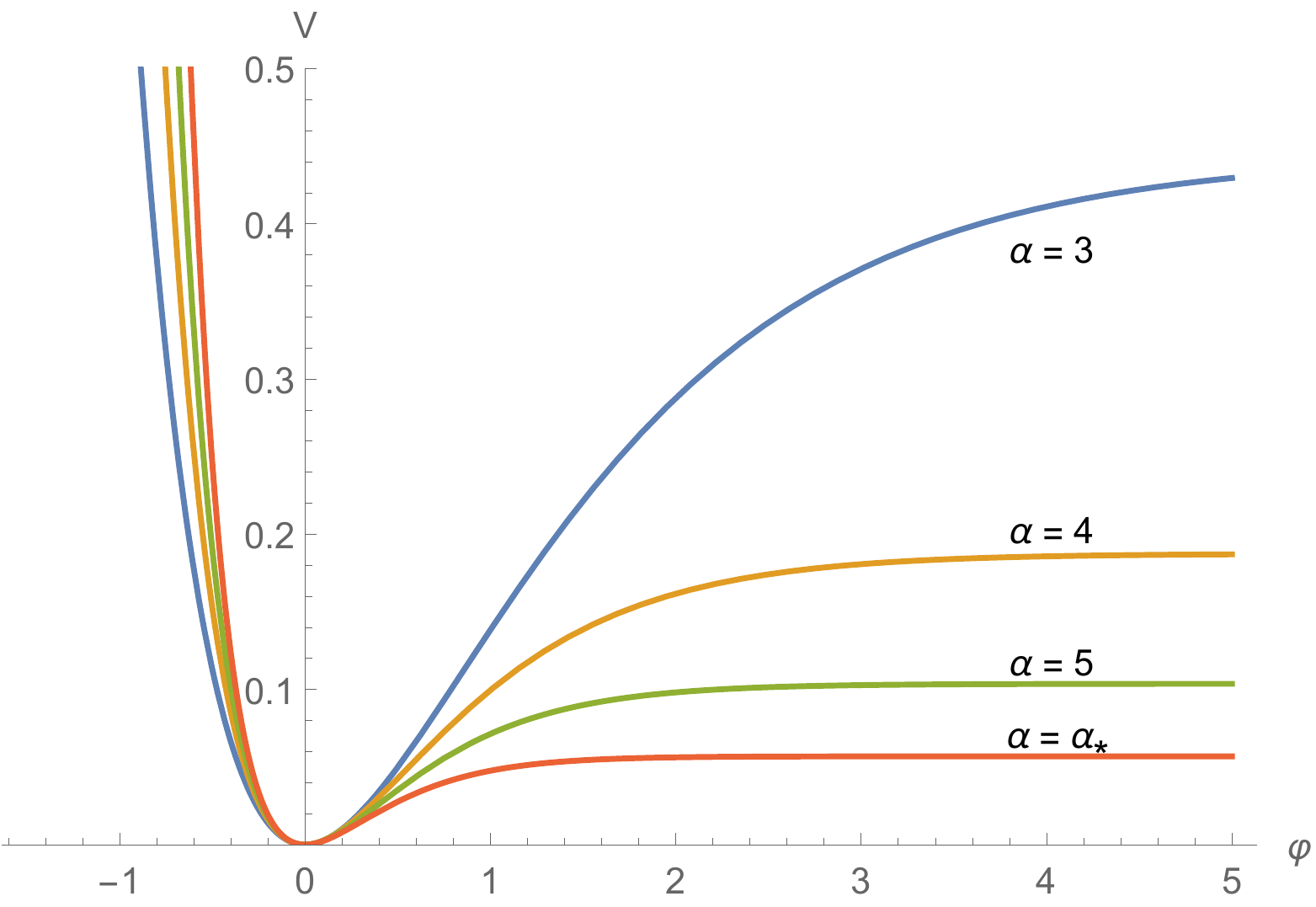}
  \caption{}
  \label{Fig_V_general1}
\end{subfigure}
\hspace{1em}
\begin{subfigure}{.47\textwidth}
  \centering
  \includegraphics[width=1\linewidth]{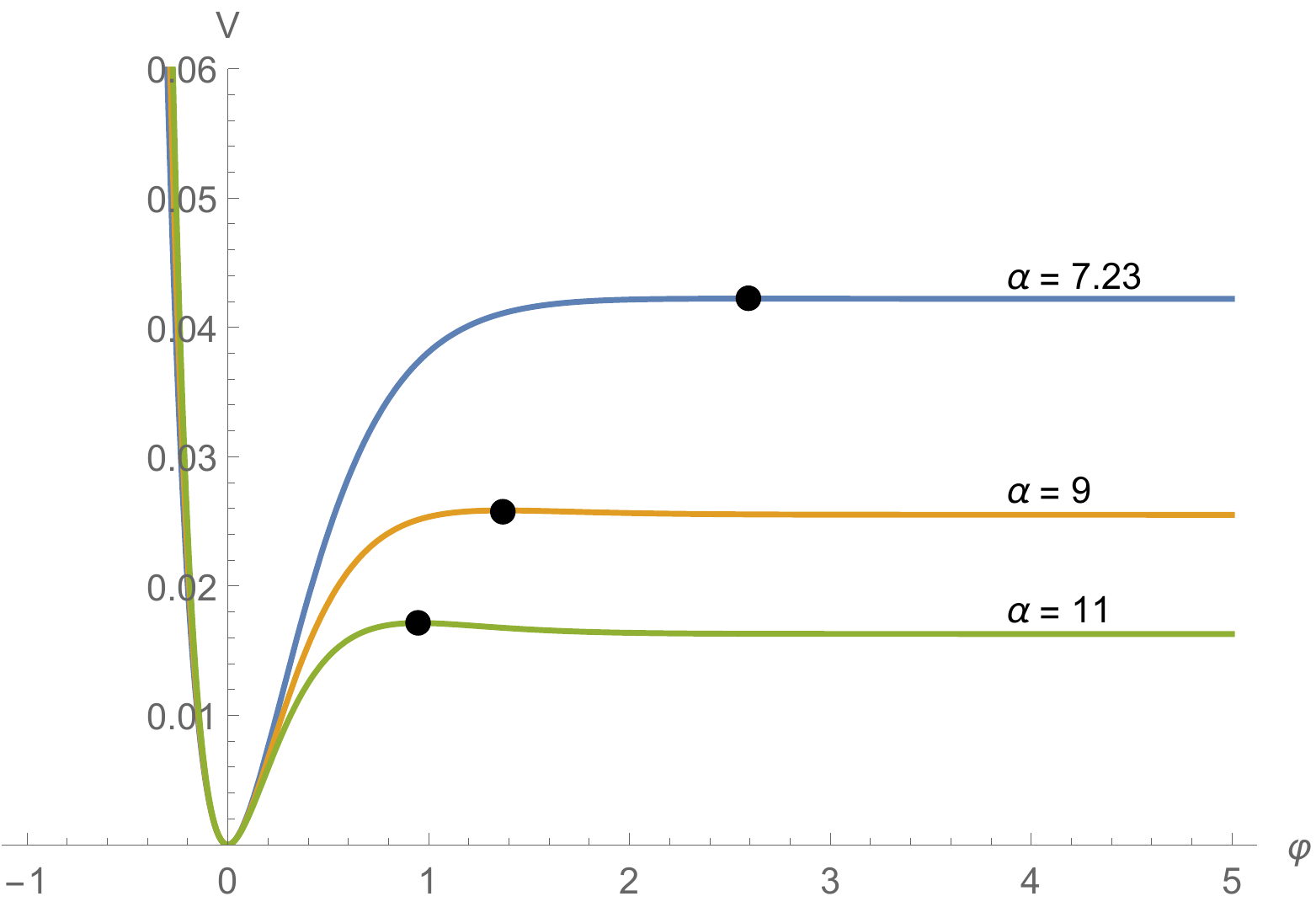}
  \caption{}
  \label{Fig_V_general2}
\end{subfigure}
\captionsetup{width=1\linewidth}
\caption{The scalar potential \eqref{V_I}\eqref{V_II} at $\tau=0$ for $\mu=2$, $\lambda=1$, $V_0=0$ and different values of $\alpha$. Subfigure (a) includes some Starobinsky-like examples, including the marginal case $\alpha_*\approx 6.37$. Subfigure (b) includes hilltop examples where the marker points represent local maxima.}
\label{Fig_V_general}
\end{figure}

When $\alpha>\alpha_*$ the potential develops a local maximum at $t_{0(2)}$ given by Eq. \eqref{crit_general}, and thus does not belong to Starobinsky-type models. However, viable (hilltop) inflation is still possible as confirmed in Ref. \cite{Aldabergenov:2019aag} where the analyzed models include similar scalar potential. In that work it is found that the spectral tilt compatible with PLANCK data \cite{Akrami:2018odb} can be reproduced by the model as long as $\alpha\lesssim 7.23$. This result applies here as well. The plots of the scalar potential at $\tau=0$ for different values of $\alpha$ are given in Figure \ref{Fig_V_general}. When $\alpha>7.23$, the curvature of the potential on the left-side of the maxima is too large to accommodate the observed value of the tilt $n_s$, as demonstrated by the examples of Figure \ref{Fig_V_general2}.

At the minimum $\varphi=0$ or $t=t_0$, the inflaton F-term vanishes, while
\begin{equation}
    \langle \tilde{F}^s\rangle=-\frac{\gamma}{2\alpha}t_0^{\frac{\alpha}{2}-1}=-\sqrt{\frac{3}{\alpha^3}}\mu~,
\end{equation}
where we used Eq. \eqref{gamma_V0} with $V_0=0$. The gravitino mass reads
\begin{equation}
    m_{3/2}^2=\frac{4(\alpha-2)^{\alpha-2}\mu^\alpha}{(2\alpha)^\alpha\lambda^{\alpha-2}}~.\label{m32}
\end{equation}

\subsection{The case II: \texorpdfstring{$\gamma=0$}{Lg} and \texorpdfstring{$n=\alpha$}{Lg}}

Upon fixing $\gamma=0$ and $n=\alpha$, the K\"ahler potential and superpotential take the form
\begin{gather}
    K=-\alpha\log\left(T+\overbar{T}-\frac{S\overbar{S}}{(T+\overbar{T})^{\alpha-1}}\right)~,\label{K_II}\\
    W=\lambda-\mu T+\beta S~.\label{W_II}
\end{gather}
Here $\beta$ can be eliminated via
\begin{equation}
    \frac{\beta^2}{\alpha}=V_0+\frac{12(\alpha-2)^{\alpha-2}\mu^\alpha}{(2\alpha)^\alpha\lambda^{\alpha-2}}~,\label{beta_V0}
\end{equation}
and the potential takes the form
\begin{multline}
    V=V_0+\frac{(\alpha-2)^{\alpha-2}\mu^\alpha}{(2\alpha)^\alpha\lambda^{\alpha-2}}\left\lbrace12+\alpha(\alpha^2-7\alpha+4)e^{(2-\alpha)\sqrt{\frac{2}{\alpha}}\varphi}+2\alpha(\alpha-2)(5-\alpha)e^{(1-\alpha)\sqrt{\frac{2}{\alpha}}\varphi}+\right.\\\left.+(\alpha-2)^2(\alpha-3)e^{-\alpha\sqrt{\frac{2}{\alpha}}\varphi}+2\alpha(\alpha-3)e^{-\alpha\sqrt{\frac{2}{\alpha}}\varphi}\tau^2\right\rbrace~,\label{V_II}
\end{multline}
Setting $\alpha=3$ leads to the potential \eqref{V_3II} with vanishing sinflaton mass. For $\alpha>3$, however, the mass term for $\tau$ is generated. The only difference between Eqs. \eqref{V_I} and \eqref{V_II} is the presence of the second term in the square brackets of Eq. \eqref{V_I} that prevents the vanishing of the sinflaton mass for $\alpha=3$ and can be traced back to the $ST$ coupling in the superpotential \eqref{W_I}. The potential \eqref{V_II} is exactly the same as the one described in Ref. \cite{Aldabergenov:2019aag} (see the case $\omega_1<0$ there). However, in contrast with the models described here, in Ref. \cite{Aldabergenov:2019aag} we used alternative Fayet--Iliopoulos D-terms \cite{Cribiori:2017laj,Kuzenko:2018jlz} to generate constant contribution to the scalar potential, whereas here the constant term is obtained from the $S$- or $ST$-term in the superpotential, while the nilpotency of $\mathbf{S}$ plays a crucial role.

As regards the F-terms,
\begin{equation}
    \langle \tilde{F}^s\rangle=-\frac{\beta}{\alpha}t_0^{\alpha/2}=-\frac{2\sqrt{3}}{\sqrt{\alpha}(\alpha-2)}\lambda~,
\end{equation}
while $\langle \tilde{F}^t\rangle$ once again vanishes. The gravitino mass is given by Eq. \eqref{m32}.

For the potentials \eqref{V_I} and \eqref{V_II} at $\tau=0$ and $\varphi\gg 1$ (slow-roll), the Hubble parameter is given by
\begin{equation}
    H^2\simeq \frac{4(\alpha-2)^{\alpha-2}\mu^\alpha}{(2\alpha)^\alpha\lambda^{\alpha-2}}=m^2_{3/2}~,\label{Hubble_general}
\end{equation}
and the observed scalar amplitude fixes the parameter ratio $\mu^\alpha/\lambda^{\alpha-2}$ at $\sim 10^{-8}$ or $10^{-7}$, depending on the exact value of $\alpha$.

\section{Full component action in unitary gauge}

We derive here the full component action including fermions, for the both cases (I and II). Once the nilpotency constraint $\mathbf{S}^2=0$ is solved as $S=(\chi^s)^2/(2F^s)$, the goldstino sector will be generated where supersymmetry is non-linearly realized. But local supersymmetry allows us to choose the gauge where $\chi^s=0$ (unitary gauge) that greatly simplifies the action. After proper rescaling of the inflatino, $\chi\rightarrow\chi t_0/\sqrt{\alpha}$ (we can drop the upper index $t$ of $\chi^t$), the full Lagrangian reads
\begin{multline}
    e^{-1}{\cal L}=\frac{1}{2}R-\frac{1}{2}(\partial_m\varphi)^2-\frac{1}{2}e^{-2\sqrt{\frac{2}{\alpha}}\varphi}(\partial_m\tau)^2-\epsilon^{klmn}\overbar{\psi}_k\overbar{\sigma}_lD_m\psi_n-ie^{-2\sqrt{\frac{2}{\alpha}}\varphi}\overbar{\chi}\overbar{\sigma}^mD_m\chi-\\
    -\left[\frac{1}{2}e^{-2\sqrt{\frac{2}{\alpha}}\varphi}\left(e^{\sqrt{\frac{2}{\alpha}}\varphi}\partial_m\varphi-i\partial_m\tau\right)\chi\sigma^n\overbar{\sigma}^m\psi_n+{\rm h.c.}\right]+\\
    +\frac{1}{4}e^{-2\sqrt{\frac{2}{\alpha}}\varphi}\left(i\epsilon^{klmn}\psi_k\sigma_l\overbar{\psi}_m+\psi_m\sigma^n\overbar{\psi}^m\right)\chi\sigma_n\overbar{\chi}-\frac{\alpha-4}{8\alpha}e^{-4\sqrt{\frac{2}{\alpha}}\varphi}\chi^2\overbar{\chi}^2-\\
    -\left[\frac{(\alpha-2)^\alpha\mu^\alpha}{(2\alpha)^\alpha\lambda^{\alpha-2}}\right]^{\frac{1}{2}}e^{-\sqrt{\frac{2}{\alpha}}\varphi}\left\lbrace \left(1-\frac{\alpha}{\alpha-2}e^{\sqrt{\frac{2}{\alpha}}\varphi}+i\frac{\sqrt{2\alpha}}{\alpha-2}\tau\right)\psi_m\sigma^{mn}\psi_n+\right.\\
    \left.+\left[i\sqrt{\frac{\alpha}{2}}\left(1-e^{-\sqrt{\frac{2}{\alpha}}\varphi}\right)-\frac{\alpha}{\alpha-2}e^{-\sqrt{\frac{2}{\alpha}}\varphi}\tau\right]\chi\sigma^m\overbar{\psi}_m+\right.\\
    \left.+\frac{\alpha-1}{2}e^{-2\sqrt{\frac{2}{\alpha}}\varphi}\left(1-\frac{\alpha-4}{\alpha-2}e^{\sqrt{\frac{2}{\alpha}}\varphi}-i\frac{\sqrt{2\alpha}}{\alpha-2}\tau\right)\chi^2+{\rm h.c.} \right\rbrace-V~,\label{L_full}
\end{multline}
where spinor indices are suppressed, and the combined Lorentz-/K\"ahler-covariant derivatives of the fermions are
\begin{gather}
    D_m\psi_n\equiv\partial_m\psi_n+\psi_n\omega_m+\frac{1}{4}(K_T\partial_m T-K_{\overbar{T}}\partial_m\overbar{T})\psi_n=\partial_m\psi_n+\psi_n\omega_m-i\sqrt{\frac{\alpha}{8}}e^{-\sqrt{\frac{2}{\alpha}}\varphi}\partial_m\tau\psi_n~,\\
    D_m\chi\equiv\partial_m\chi+\chi\omega_m+\Gamma^{T}_{TT}\partial_mT\chi-\frac{1}{4}(K_T\partial_m T-K_{\overbar{T}}\partial_m\overbar{T})\chi=\\
    =\partial_m\chi+\chi\omega_m-\sqrt{\frac{2}{\alpha}}\partial_m\varphi\chi+i\frac{\alpha-4}{2\sqrt{2\alpha}}e^{-\sqrt{\frac{2}{\alpha}}}\partial_m\tau\chi~.
\end{gather}

The first line in Eq. \eqref{L_full} represents the kinetic terms, while the second line represents the coupling between $\chi$, $\psi_m$, and derivatives of the scalars. Four-fermion interactions are included in the third line, and the last three lines consist of fermion mass terms as well as the scalar potential $V$ which is the only difference between the models I and II: for the case I $V$ is given by Eq. \eqref{V_I}, and for the case II by Eq. \eqref{V_II}.

The $\varphi$-, $\psi_m$-, and $\chi$-masses (around $\varphi=0$) are the same between models I and II,
\begin{gather}
    m_{\varphi}=2\left[\frac{(\alpha+1)(\alpha-2)^{\alpha-1}\mu^\alpha}{(2\alpha)^\alpha\lambda^{\alpha-2}}\right]^{\frac{1}{2}}~,\nonumber\\
    m_{3/2}=2\left[\frac{(\alpha-2)^{\alpha-2}\mu^\alpha}{(2\alpha)^\alpha\lambda^{\alpha-2}}\right]^{\frac{1}{2}}~,~~~m_{\chi}=2\frac{\alpha-1}{\alpha-2}\left[\frac{(\alpha-2)^{\alpha-2}\mu^\alpha}{(2\alpha)^\alpha\lambda^{\alpha-2}}\right]^{\frac{1}{2}}~,
\end{gather}
whereas the $\tau$-mass is different,
\begin{equation}
    m^{\rm I}_\tau=2\left[[12+\alpha^2(\alpha-3)]\frac{(\alpha-2)^{\alpha-2}\mu^\alpha}{\alpha(2\alpha)^\alpha\lambda^{\alpha-2}}\right]^{\frac{1}{2}}~,~~~ m^{\rm II}_\tau=2\left[\frac{\alpha(\alpha-3)(\alpha-2)^{\alpha-2}\mu^\alpha}{(2\alpha)^\alpha\lambda^{\alpha-2}}\right]^{\frac{1}{2}}~.
\end{equation}

\begin{figure}
\centering
  \includegraphics[width=.62\linewidth]{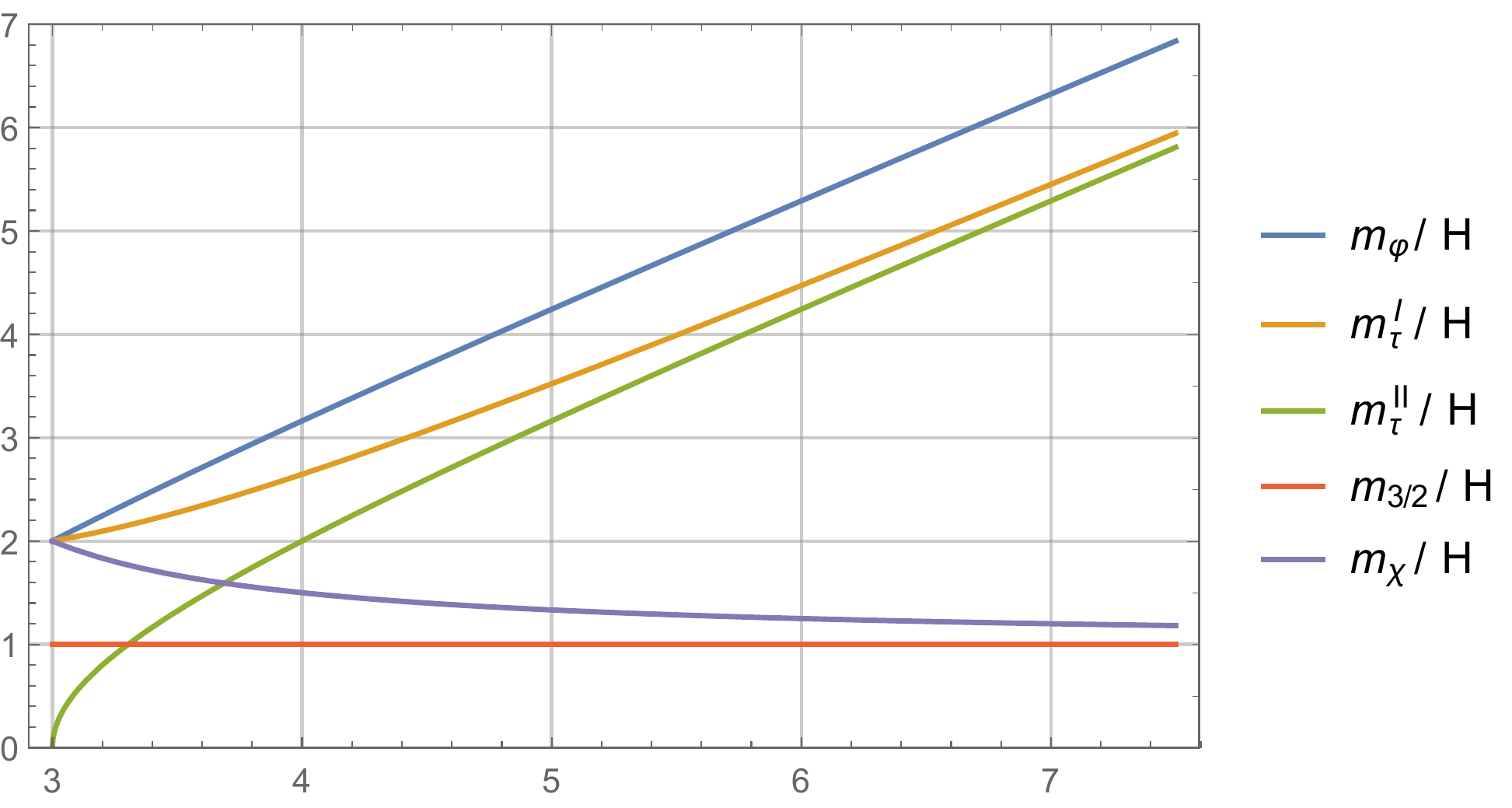}
\caption{The mass-to-Hubble ratios of the inflaton $\varphi$, sinflaton $\tau$, gravitino $\psi_m$, and inflatino $\chi$. The horizontal axis represents $\alpha$.}
\label{Fig_m}
\end{figure}

To illustrate the relation between the masses at different $\alpha$, we include Figure \ref{Fig_m} where the mass-to-Hubble ratios $m_\varphi/H$, $m^{\rm I,II}_{\tau}/H$, $m_{3/2}/H$, and $m_\chi/H$ are plotted as functions of $\alpha$ (after using the expression \eqref{Hubble_general} for the inflationary Hubble parameter, $\lambda$ and $\mu$ dependence cancels out). In the case I with $\alpha=3$, the masses of $\varphi$, $\tau$, and $\chi$ coincide and are twice the gravitino mass that is equal to the Hubble parameter. Once we depart from the Starobinsky case $\alpha=3$, the masses split: $m_\varphi$ and $m_\tau$ almost-linearly grow compared to $H$ (and $m_{3/2}$), with $\varphi$ becoming the heavier one, whereas $m_\chi$ asymptotically approaches $H$. In the case II the same is true except that $m_\tau$ is zero for $\alpha=3$, and with growing $\alpha$ it approaches the behavior of $m^{\rm I}_\tau$.

\section{Slow-roll approximation}

Let us consider the slow-roll regime of the Starobinsky-like scenario that is available for $3\leq\alpha\leq\alpha_*$, $\alpha_*=(7+\sqrt{33})/2\approx 6.37$. Assuming that $\tau$ is stabilized at $\tau=0$, the potential for the both cases I and II is given by
\begin{equation}
    V\sim 1+\frac{\alpha}{12}(\alpha^2-7\alpha+4)e^{(2-\alpha)\sqrt{\frac{2}{\alpha}}\varphi}+\ldots~,\label{V_approx}
\end{equation}
where the overall constant factor is irrelevant. We use the standard definition of the slow-roll parameters
\begin{equation}
    \epsilon_i\equiv\left.\frac{1}{2}\left(\frac{V'}{V}\right)^2\right|_{\varphi=\varphi_i}~,~~~\eta_i\equiv\left.\frac{V''}{V}\right|_{\varphi=\varphi_i}~,\label{slow_roll_param}
\end{equation}
where $\varphi_i$ is field value at the start of inflation (horizon crossing). The slow-roll parameters are then related to the observable spectral tilt and tensor-to-scalar ratio,
\begin{equation}
    n_s=1+2\eta_i-6\epsilon_i~,~~~r=16\epsilon_i~.\label{ns_r}
\end{equation}
In order to express these in terms of the elapsed number of e-foldings $N_e$, we use
\begin{equation}
    N_e=\int^{\varphi_i}_{\varphi_f}d\varphi\frac{V}{V'}~,\label{e-foldings}
\end{equation}
where $\varphi_f$ can be neglected for the approximate results.

Using the formulae \eqref{V_approx} to \eqref{e-foldings} we obtain
\begin{equation}
    n_s\simeq 1-\frac{2}{N_e}~,~~~r\simeq \frac{4\alpha}{(\alpha-2)^2N_e^2}~,\label{ns_r_result}
\end{equation}
which is the main result of this section.

One caveat here is that when $\alpha=\alpha_*$, the leading $\varphi$-term in the potential \eqref{V_approx} vanishes, and the next term should be included, i.e.,
\begin{equation}
    V\sim 1-\frac{\alpha}{6}(\alpha-2)(\alpha-5)e^{(1-\alpha)\sqrt{\frac{2}{\alpha}}\varphi}+\ldots~.
\end{equation}
In this case the tensor-to-scalar ratio is modified as
\begin{equation}
    r\simeq\frac{4\alpha}{(\alpha-1)^2N_e^2}~.
\end{equation}

Nevertheless, Eq. \eqref{ns_r_result} still provides a good approximation for our purposes. The output of Eq. \eqref{ns_r_result} can be compared with the numerical results of Ref. \cite{Aldabergenov:2019aag} (see Table 1 for $\omega_1<0$ there), because the $\varphi$-dependent scalar potential with $\omega_1<0$ in that work is identical to what we obtained here.

\section{Dual gravitational actions}

Let us first review the dual gravitational action of the ADFS model. Using the K\"ahler potential and superpotential of Eqs. \eqref{K_ADFS}\eqref{W_ADFS}, the superspace action can be explicitly written as \cite{Antoniadis:2014oya}
\begin{eqnarray}
    {\cal L}&=&\int d^2\Theta 2{\cal E}\left[\frac{3}{8}(\overbar{\cal D}^2-8{\cal R})({\bf T}+\overbar{\bf T}-{\bf S\overbar{S}})+\lambda+\beta {\bf S}+\gamma {\bf ST}\right]+{\rm h.c.}=\nonumber\\
    &=&\int d^2\Theta 2{\cal E}\left[-\frac{3}{8}(\overbar{\cal D}^2-8{\cal R}){\bf S\overbar{S}}+\lambda+\beta {\bf S}-{\bf T}(6{\cal R}-\gamma{\bf S})\right]+{\rm h.c.}~,
\end{eqnarray}
where we used the superspace identity
\begin{equation}
    \int d^2\Theta 2{\cal E}(\overbar{\cal D}^2-8{\cal R})({\bf T}+\overbar{\bf T})+{\rm h.c.}=-16\int d^2\Theta 2{\cal E}{\cal R}{\bf T}+{\rm h.c.}
\end{equation}

Varying the action with respect to $\bf T$, we obtain the relation ${\bf S}=6{\cal R}/\gamma$ so that we can eliminate $\bf S$ and arrive at the higher-derivative (gravitational) action,
\begin{equation}
    {\cal L}=\int d^2\Theta 2{\cal E}\left[\frac{6\beta}{\gamma}{\cal R}-\frac{27}{2\gamma^2}(\overbar{\cal D}^2-8{\cal R}){\cal R\overbar{R}}+\lambda\right]+{\rm h.c.}
\end{equation}
The proper normalization of the Einstein--Hilbert part (the first term) requires setting $\beta=-\gamma/2$,~\footnote{Alternatively, the constant factor $\sim \beta/\gamma$ in front of the Einstein--Hilbert term can be absorbed by Weyl-rescaling of the metric, but this is equivalent to setting $\beta=-\gamma/2$ because either way we are left with just two independent (effective) parameters.} while the nilpotency condition ${\bf S}^2=0$ translates into ${\cal R}^2=0$. The nilpotency of $\cal R$ can be included in the action by adding a Lagrange multiplier chiral superfield $\bf Z$ so that the final Lagrangian reads
\begin{equation}
    {\cal L}=\int d^2\Theta 2{\cal E}\left[-3{\cal R}-\frac{27}{2\gamma^2}(\overbar{\cal D}^2-8{\cal R}){\cal R\overbar{R}}+\lambda+{\bf Z}{\cal R}^2\right]+{\rm h.c.}~
\end{equation}

Next, let us consider the dualization of our first model given by Eqs. \eqref{K_3I}\eqref{W_3I} with $\beta=0$. Following similar steps as above we obtain
\begin{equation}
    {\cal L}=\int d^2\Theta 2{\cal E}\left[-\frac{3}{8}(\overbar{\cal D}^2-8{\cal R}){\bf S\overbar{S}}+\lambda-{\bf T}(6{\cal R}-\gamma{\bf S}+\mu)\right]+{\rm h.c.}~
\end{equation}
Varying with respect to $\bf T$ leads to the equation
\begin{equation}
    {\bf S}=\frac{6}{\gamma}\left({\cal R}+\frac{\mu}{6}\right)~,\label{SR_duality}
\end{equation}
which means that the nilpotency ${\bf S}^2=0$ corresponds to the $\cal R$-constraint
\begin{equation}
    \left({\cal R}+\frac{\mu}{6}\right)^2=0~.\label{R_constr}
\end{equation}

Eliminating $\bf S$ via \eqref{SR_duality} and adding the Lagrange multiplier $\bf Z$ for the constraint \eqref{R_constr} we arrive at the dual gravitational action,
\begin{equation}
    {\cal L}=\int d^2\Theta 2{\cal E}\left[-\frac{27}{2\gamma^2}(\overbar{\cal D}^2-8{\cal R})\left|{\cal R}+\frac{\mu}{6}\right|^2+\lambda+{\bf Z}\left({\cal R}+\frac{\mu}{6}\right)^2\right]+{\rm h.c.}~\label{L_dual}
\end{equation}
In contrast with the ADFS case, here the normalization of the Einstein--Hilbert term by constant Weyl-rescaling does not reduce the number of independent parameters.

This model has similar features to the one proposed in Ref. \cite{Hasegawa:2015era}: both models have "shifted" nilpotency constraints for the curvature superfield $\cal R$, and both models lead to Starobinsky inflation with de Sitter vacuum after inflation where supersymmetry is spontaneously broken. However, the actions are different (the difference in the K\"ahler potentials is also clear on the dual scalar-tensor side), as well as the predicted SUSY breaking scales -- the gravitino mass in \cite{Hasegawa:2015era} is of order $10^8$ GeV.

Unfortunately, the model given by Eqs. \eqref{K_3II}\eqref{W_3II} -- as well as the generalized models of Section 3 -- cannot be dualized into higher-derivative supergravities (at least not by the standard procedure that we used above).

\section{Conclusion}

In this work we introduced alternative models of Volkov--Akulov--Starobinsky supergravity building upon the ADFS model \cite{Antoniadis:2014oya}. In the ADFS model, after inflation the vacuum value of the auxiliary component of the goldstino superfield vanishes, rendering the solution to the nilpotency constraint singular. We studied two different types of modifications to the ADFS setup that can improve the vacuum structure of the F-terms as
\begin{equation}
    \langle F^t\rangle=0~,~~~\langle F^s\rangle\neq 0~,
\end{equation}
while preserving the no-scale-type K\"ahler potential.

Moreover, we showed that the K\"ahler potential can be generalized while keeping all the desired properties, as
\begin{equation}
    K=-\alpha\log\left(T+\overbar{T}-\frac{S\overbar{S}}{(T+\overbar{T})^{n-1}}\right)~.\label{K_conc}
\end{equation}
For the superpotential
\begin{equation}
    W=\lambda-\mu T+\beta S+\gamma ST~,~~~\lambda\mu>0~,\label{W_conc}
\end{equation}
Starobinsky-like inflation with de Sitter vacuum (after inflation) is possible for $3\leq\alpha\leq\alpha_*$ ($\alpha_*=(7+\sqrt{33})/2$) and hilltop inflation that agrees with CMB data \cite{Akrami:2018odb} is possible for $\alpha_*<\alpha\lesssim 7.23$, if we choose $\{\beta=0,~n=\alpha-2\}$ or $\{\gamma=0,~n=\alpha\}$. We found that the scalar potential in these two cases is very similar to the one described in Ref. \cite{Aldabergenov:2019aag}: the potential \eqref{V_II} of model II exactly coincides with the potential of \cite{Aldabergenov:2019aag}, while the potential \eqref{V_I} of model I has a different $\tau^2$-term with larger $m_\tau$ (see e.g. Figure \ref{Fig_m}). Also, in Ref. \cite{Ellgan:2019zze} two-field analysis was performed for the same class of models as in \cite{Aldabergenov:2019aag}, where isocurvature effects are shown to be small. This implies that in model I isocurvature effects should be even more suppressed compared to model II, due to the larger $\tau$-mass, and substantially larger effective $\tau$-mass for $\varphi\gg 1$.

We derived the full component action for the general setup \eqref{K_conc}, \eqref{W_conc}, and showed the behavior of the mass spectrum at different $\alpha$. With the exception of $\alpha=n=3$ with $\gamma=0$ where the sinflaton mass vanishes, all the fields generally have large masses comparable to the inflationary Hubble scale, while $\langle F^s\rangle$ is not fixed by CMB observations.

Slow-roll approximation can be used when $3\leq\alpha\leq\alpha_*$, and is shown to lead to the prediction
\begin{equation}
    n_s\simeq 1-\frac{2}{N_e}~,~~~r\simeq\frac{4\alpha}{(\alpha-2)^2N_e^2}~.\label{nsr_conc}
\end{equation}
Comparing these predictions with the numerical results of \cite{Aldabergenov:2019aag}, it can be seen that even for $\alpha_*<\alpha\lesssim 7.23$ (hilltop case) Eq. \eqref{nsr_conc} provides good estimates.

Finally, we derived the gravitational dual action of the model \eqref{K_3I}\eqref{W_3I}, and showed that the nilpotency constraint on the scalar-tensor side, ${\bf S}^2=0$, is translated into the "shifted" nilpotency constraint for the chiral curvature superfield, $\left({\cal R}+\mu/6\right)^2=0$ (in comparison, in the gravitational ADFS model the curvature superfield satisfies ${\cal R}^2=0$). The rest of the models that we proposed cannot be dualized into higher-derivative SUGRA by the standard procedure due to the forms of the corresponding K\"ahler potentials.

\section*{Acknowledgements}

The author was supported by the CUniverse research promotion project of Chulalongkorn University under the grant reference CUAASC, and the Ministry of Education and Science of the Republic of Kazakhstan under the grant reference BR05236322.

\section*{Appendix}

We follow the notations and conventions of Ref. \cite{Wess:1992cp}, where the superspace action for the chiral superfield coupled to standard Poincar\'e supergravity reads ($M_P=1$ and "mostly plus" metric signature is used)
\begin{equation}
    {\cal L}=\int d^2\Theta 2{\cal E}\left[\frac{3}{8}(\overbar{\cal D}^2-8{\cal R})e^{-K(\mathbf{\Phi}^i,\overbar{\mathbf{\Phi}}^i)/3}+W(\mathbf{\Phi}^i)\right]+{\rm h.c.}~,\label{App_L_super}
\end{equation}
where $\cal E$ is the chiral density superfield, $\cal R$ is the chiral curvature superfield, ${\cal D}_\alpha,\overbar{\cal D}_{\dot{\alpha}}$ are the superspace (fermionic) covariant derivatives with ${\cal D}^2\equiv{\cal D}^\alpha{\cal D}_\alpha$ and $\overbar{\cal D}^2\equiv\overbar{\cal D}_{\dot{\alpha}}\overbar{\cal D}^{\dot{\alpha}}$, and $K$ and $W$ are function of a given set of chiral superfields $\mathbf{\Phi}^i$. The operator $(\overbar{\cal D}^2-8{\cal R})$ is the chiral projector in curved superspace, so that the first term in Eq. \eqref{App_L_super} is D-term.

The component expansion of $\cal E$ and $\cal R$ is given by
\begin{eqnarray}
    2{\cal E}&=&e\left[1+i\Theta\sigma^m\overbar{\psi}_m-\Theta^2(\overbar{M}+\overbar{\psi}_m\overbar{\sigma}^{mn}\overbar{\psi}_n)\right]~,\\
    {\cal R}&=&-\frac{1}{6}\bigg[M+\Theta(\sigma^m\overbar{\sigma}^n\psi_{mn}-i\sigma^m\overbar{\psi}_mM+i\psi_mb^m)+.\nonumber\\
    &~&+\Theta^2\left(\frac{1}{2}R+i\overbar{\psi}^m\overbar{\sigma}^n\psi_{mn}+\frac{2}{3}M\overbar{M}+\frac{1}{3}b_mb^m-i\nabla_mb^m+\right.\nonumber\\
    &~&+\left.\frac{1}{2}\overbar{\psi}_m\overbar{\psi}^mM-\frac{1}{2}\psi_m\sigma^m\overbar{\psi}_nb^n+\frac{1}{8}\varepsilon^{abcd}(\overbar{\psi}_a\overbar{\sigma}_b\psi_{cd}+\psi_a\sigma_b\overbar{\psi}_{cd})\right)\bigg]~,\label{R_expansion}
\end{eqnarray}
where $e\equiv{\rm det}(e^a_m)$ is determinant of the frame field ($a$ -- Lorentz index, $m$ -- Einstein index), and $\psi_{mn}\equiv\tilde{D}_m\psi_n-\tilde{D}_n\psi_m$ with $\tilde{D}_m\psi_n\equiv\partial_m\psi_n+\psi_n\omega_m$ (Lorentz-covariant derivative). Spinor indices are suppressed. The vector $b_m$ and complex scalar $M$ represent the old-minimal set of SUGRA auxiliary fields. These fields become dynamical when the superspace action is extended by (gravitational) higher-derivative terms (see e.g. \cite{Ketov:2013dfa} for more details). We use the definition of the scalar curvature $R$ that has the opposite sign compared to Ref. \cite{Wess:1992cp}.

The generic matter chiral superfield has the standard expansion,
\begin{equation}
    {\bf\Phi}^i=\Phi^i+\sqrt{2}\Theta\chi^{i}+\Theta^2F^{i}~.\label{Phi_expansion}
\end{equation}

After expanding the Lagrangian \eqref{App_L_super} in terms of the component fields, eliminating the auxiliary components, and Weyl-rescaling to Einstein frame, we obtain the scalar potential
\begin{equation}
    V=e^K\left(K^{i\bar{j}}D_iW D_{\bar{j}}\overbar{W}-3|W|^2\right)~,\label{V_SUGRA}
\end{equation}
where $K=K(\Phi^i,\overbar{\Phi}^i)$ is the component K\"ahler potential, $W=W(\Phi^i)$ is the component superpotential and the following standard notation is used
\begin{gather}
    K_{i\bar{j}}\equiv\frac{\partial^2 K}{\partial\Phi^i\partial\overbar{\Phi}^j}~,~~~K^{i\bar{j}}\equiv K^{-1}_{i\bar{j}}~,~~~D_iW\equiv\frac{\partial W}{\partial\Phi^i}+W\frac{\partial K}{\partial\Phi^i}~.
\end{gather}
$D_iW$ are proportional to the corresponding auxiliary F-terms via their algebraic equations of motion,
\begin{equation}
    F^i=-e^{K/3}K^{i\bar{j}}D_{\bar{j}}\overbar{W}~,~~~\overbar{F}^j=-e^{K/3}K^{i\bar{j}}D_iW~.\label{F_DW}
\end{equation}
There is a difference between the Wess--Bagger definition of the auxiliary field $F^i$, as in Eqs. \eqref{Phi_expansion}\eqref{F_DW}, and a more common definition
\begin{equation}
    \tilde{F}^i=-e^{K/2}K^{i\bar{j}}D_{\bar{j}}\overbar{W}~,~~~\overbar{\tilde{F}}^j=-e^{K/2}K^{i\bar{j}}D_iW~.\label{tilde_F_DW}
\end{equation}
The latter is motivated by the fact that the scalar potential can be written as
\begin{equation}
    V=K_{i\bar{j}}\tilde{F}^i\overbar{\tilde{F}}^j+\ldots~,
\end{equation}
whereas if we use $F^i$, an extra $K$-dependent factor will appear,
\begin{equation}
    V=e^{K/3}K_{i\bar{j}}F^i\overbar{F}^j+\ldots~.
\end{equation}
The two fields are related by $F^i=e^{-K/6}\tilde{F}^i$.

\bibliography{Bibliography.bib}{}
\bibliographystyle{utphys.bst}

\end{document}